\documentclass[11pt,letterpaper]{article} 
\newcommand{\bea}{\begin{eqnarray}} 
\newcommand{\eea}{\end{eqnarray}}

\newcommand{\eins}{{\mbox{\sf 1\hspace{-3pt}I\hspace{-4pt}\_}}}

\newcommand{\bq}{\begin{equation}}
\newcommand{\eq}{\end{equation}}
\newcommand{\bgar}{\begin{array}}
\newcommand{\ear}{\end{array}}

\renewcommand{\rho}{\varrho}

\renewcommand{\d}{\partial}

\newcommand{\tr}{{\mathrm {tr}}}
\newcommand{\bi}{{\mathbf i}}
\newcommand{\bk}{{\mathbf k}}
\newcommand{\U}{{\mathrm U}}
\newcommand{\SU}{{\mathrm {SU}}}
\renewcommand{\L}{{\mathcal L}}

\begin{document}
\begin{center}
\huge{\bf{\mbox{Gauge Couplings and Group Dimensions} \\
in the Standard Model}}
\end{center}

\vspace{0.2cm}

\begin{center}
\Large{Ralf Alfons Engeldinger\footnote{Current mail address: Lyman Laboratory of Physics, Harvard University, Cambridge, MA 02138}}
\end{center}

\begin{center}
\small {oatesian@feynman.harvard.edu}
\end{center}

\vspace{3cm}

\begin{abstract}
The gauge field term in the Standard Model Lagrangian is slightly rewritten, suggesting that the three gauge couplings have absorbed factors which depend on the dimensions of the corresponding gauge groups.  The ratios of the physical couplings may turn out to be dominated by these factors, with deviations due to quantum corrections.
\end{abstract}

\renewcommand{\arraystretch}{1.7}
\section{\normalsize Introduction}
A slight modification of the formulation of the Lagrangian of the Standard Model implies that the coupling $g_\bk$ of a simple gauge group of dimension $\;\dim_\bk\;$ absorbs a hidden factor proportional to $\sqrt{\dim_\bk}$. This suggests the possibility that these dimension-dependent factors are the ultimate  origin of the experimentally observed differences of the 
coupling strengths of the three gauge interactions. We state the hypothesis, that the ratios $\displaystyle \frac{g_\bk}{\sqrt{\dim_\bk}}$ are equal for $\SU(3)$, $\SU(2)$ and $\U(1)$ before quantum corrections. This then implies:
\begin{itemize}
\item $\sin^2\vartheta_{W\mathrm{cl}} = 0.25$ before quantum corrections 
\item Strong coupling of $\SU(3)$ gauge fields as compared to electroweak: \\ \mbox{$\displaystyle\frac{\alpha_{s\;\mathrm{cl}}}
{\alpha_{\mathrm{cl}}}=\frac{32}3\approx 10.67$} 
\item Unification of the gauge couplings {\em without} unification of the gauge group: A larger, simple gauge group is not required, nor are, therefore, unobserved heavy gauge bosons.
\end{itemize}

\section{\normalsize Action Coherence Assumption}

A Yang-Mills Lagrangian of the form 
\bq\L = \L_0 - \frac1{2g^2}\tr(F^{\mu\nu}F_{\mu\nu})\eq 
can be rewritten, using the unit matrix $\eins$ in the space of the generators $T^a$ of the adjoint representation of the gauge group and 

\bq \dim = \tr\, \eins = \delta^{aa}, \eq

as

\bq \L = \frac1\dim\tr(\L_0\eins) - \frac1{2g^2}\tr(F^{\mu\nu}F_{\mu\nu})
= \frac1\dim\tr\left(\L_0\eins - \frac{\dim}{2g^2} F^{\mu\nu}F_{\mu\nu}\right). \eq

Similarly, 
\bq\L = \L_0 - \sum\limits_\bk\frac1{2g_\bk^2}
\tr_\bk(F_\bk^{\mu\nu}F_{\bk\mu\nu})\eq
can be rewritten as (omitting the $\eins_\bk$'s)

\renewcommand{\arraystretch}{2.6}

\bq \bgar{l@{{}={}}l}
\L & \displaystyle \left(\prod\limits_\bi\frac{1}{\dim_\bi}\tr_\bi\right) \left(\L_0 -\sum\limits_\bk\frac{\dim_\bk}{2g_\bk^2}
F_\bk^{\mu\nu}F_{\bk\mu\nu}\right) \\
& \displaystyle\left(\prod\limits_\bi\tr_\bi\right) \left(\frac1{\mathbf d}\L_0 -\sum\limits_\bk\frac{\dim_\bk}{2\mathbf dg_\bk^2}
F_\bk^{\mu\nu}F_{\bk\mu\nu}\right), \\ \ear\eq

where 

\bq \mathbf d=\prod\limits_\bi\dim_\bi, \eq 

and, if we include fermions,

\bq\bgar{l@{{}={}}l}
\cal L & \displaystyle\bar\psi i\gamma^\mu D_\mu\psi
-\sum\limits_\bk\frac1{2g_\bk^2}\tr (F_\bk^{\mu\nu}F_{\bk\mu\nu}) \\
& \displaystyle \left(\prod\limits_\bi\tr_\bi\right)\left(\frac1{\mathbf d}\bar\psi i\gamma^\mu\d_\mu\psi
-\sum\limits_\bk\frac{\dim_\bk}{\mathbf dg_\bk^2}
(2J_\bk^{\mu}A_{\bk\mu}
+\frac12 F_\bk^{\mu\nu}F_{\bk\mu\nu})\right) \\ \ear \eq
\renewcommand{\arraystretch}{1.7}

On the level of the matrix space, ``before'' traces, the place of the couplings $g_\bk$ is taken by the quantities \mbox{$\;\displaystyle\tilde g_\bk=\sqrt{\frac{\mathbf d}{\dim_\bk}}g_\bk,$} which we call {\em reference couplings}. It is essential, that all quantities here are normalized in such a way that the couplings $g_\bk$ don't explicitly appear in their definitions and the Yang-Mills equations. \bq \bgar{c}
D_\mu = \d_\mu-iA_\mu = \d_\mu-iT^aA^a_\mu \\
F_{\mu\nu} = i[D_\mu,D_\nu] = \d_\mu A_\nu-\d_\nu A_\mu-i[A_\mu,A_\nu] \\
J^\nu  = [D_\mu,F^{\mu\nu}], \\
\ear\eq

Especially the customary appearance of $g$ in the inhomogeneous Yang-Mills equation usually conceals the homogeneous $g$-dependence of the combined gauge field and current term in the Lagrangian which will be crucial in the following.  Since this also plays a pivotal r\^ole for the Slavnov-Taylor identities, we can expect renormalization not to invalidate the foundation of our argument.

Even though we only modify the notation of the standard Lagrangian, it is possible to test whether our reformulation has any physical content.  Relative to our version the three Standard Model couplings $g_\bk$ have absorbed the dimension-dependent factors $\displaystyle\sqrt{\frac{\dim_\bk}{\mathbf d}}$. If experimental evidence for such factors can be established then our reformulation would effectively be a correction to an oversight in the standard formulation. To secure gauge invariance of separate, individual terms is necessary but not sufficient. The normalization of separate terms needs to take into account the existence of every gauge group, even for terms on which it does not act.

{\bf Action Coherence Assumption}: {\em Lagrangian terms which are {\em not} the trace over the adjoint representation of a gauge group are to be considered as arising from the trace of the corresponding unit matrix. As a consequence, they
pick up a factor \mbox{$\dim_\bk = \tr_\bk\,\eins_\bk = \delta^{a_\bk a_\bk}$}. To compensate the effect on the invariant part, the Lagrangian as a whole is rescaled by a factor $\displaystyle\frac1{\mathbf d}$. As a consequence the gauge couplings $g_\bk$ have absorbed dimension dependent factors $\displaystyle\sqrt{\frac{\dim_\bk}{\mathbf d}}$.}

\section{\normalsize Standard Model couplings}
A correlation exists between the strengths of the three Standard Model gauge couplings (at physically relevant energy scales)
\[ g_1 < g_2 < g_3 \]
and the dimensions of the corresponding gauge groups
\[ \dim\U(1) < \dim\SU(2) < \dim\SU(3). \]

To see whether a quantitative connection between the couplings and the group dimensions can be established, we compare the ratios of couplings with those of the corresponding reference couplings.

\pagebreak

\bq\bgar{c} 
\dim_1=\dim\U(1)=1 \\
\dim_2=\dim\SU(2)=3 \\
\dim_3=\dim\SU(3)=8 \\
\mathbf d=24 
\ear\eq

\bq\bgar{c} 
\tilde g_1=\sqrt{24}\,g_1 \\
\tilde g_2=\sqrt{8}\,g_2 \\
\tilde g_3=\sqrt{3}\,g_3 \\
\ear\eq

\subsection{$\sin^2\vartheta_W$}
From the tree level definition 
$\;\displaystyle\tan\vartheta_W=\frac{g_1}{g_2}\;$ and the experimentally determined value $\sin^2\vartheta_W\approx 0.23$ we have
\bq
4.35\approx \frac1{0.23} \approx \frac1{\sin^2\vartheta_W} =
1+\frac{g_2^2}{g_1^2} = 1+\frac{3\tilde g_2^2}{\tilde g_1^2} \eq

\bq 
\frac{\tilde g_2^2}{\tilde g_1^2}\approx 1.12\qquad\mbox{vs.}
\qquad\frac{g_2^2}{g_1^2}\approx 3.35
\eq

\subsection{$\alpha_s/\alpha$}

Using the value $\alpha_s(M_Z) \approx 0.12$ for the QCD coupling in the $\overline{\mathrm {MS}}$ renormalization scheme and the low energy limit $\alpha \approx \frac1{137}$ for the fine structure constant for a qualitative approximation (using $\alpha(M_Z) \approx \frac1{129}$ is equivalent for our purposes), 
we have

\renewcommand{\arraystretch}{2.2}
\bq\bgar{l}\displaystyle
16.44 \approx \frac{\alpha_s}\alpha = \frac{g_3^2}{e^2}
= \left(\frac{1}{g_1^2}+\frac1{g_2^2}\right)g_3^2 
= \left(1+\frac{g_2^2}{g_1^2}\right)
\frac{g_3^2}{g_2^2} \approx 4.35\times
\frac{8\tilde g_3^2}{3\tilde g_2^2} \\
\displaystyle\quad\approx 11.59 \times
\frac{\tilde g_3^2}{\tilde g_2^2}
\ear\eq

\renewcommand{\arraystretch}{1.7}

\bq
\frac{\tilde g_3^2}{\tilde g_2^2} \approx 1.42\qquad
\mbox{vs.}\qquad
\frac{g_3^2}{g_2^2}\approx 3.78
\eq

\bq
\frac{\tilde g_3^2}{\tilde g_1^2} \approx 1.58
\qquad\mbox{vs.}\qquad
\frac{g_3^2}{g_1^2}\approx 12.66
\eq

\section{\normalsize Dimension-Coupling Hypothesis}
These calculations are only meant to give order-of-magnitude approximations for the ratios of the $\tilde g_\bk$'s.  The quantities $\sin^2\vartheta_W$, $\alpha_s$ and $\alpha$
are energy scale and renormalization scheme dependent, and their experimentally measured values depend on the specific processes which are used for the measurement, while the relations we use are valid on tree level.  Thus, the above ratios implicitly contain quantum corrections.  The main point is the juxtaposition of the respective 
$\tilde g_\bk$- and 
$g_\bk$-ratios which allow a reasonably valid comparison 
since they involve exactly the same approximations. We note the following: Unlike in the case of the $g_\bk$'s, the $\tilde g_\bk$-ratios are sufficiently close to $1$ to allow the possibility that {\em their deviation from $1$ is exclusively due to quantum corrections}, while {\em classically} (corresponding to tree level in perturbation theory) {\em the three $\tilde g_\bk$'s coincide.}

{\bf Dimension-Coupling Hypothesis} {\em We assume that the classical reference couplings of the gauge fields of the Standard Model gauge group} \linebreak
\mbox{SU(3)$\times$SU(2)$\times$U(1)}
{\em coincide: 
\bq\tilde g_\mathrm{cl}=\tilde g_{1\mathrm{cl}}= \tilde g_{2\mathrm{cl}} =\tilde g_{3\mathrm{cl}}.\eq 
For the classical couplings we therefore have:
\bq\bgar{c} 
\displaystyle g_{1\mathrm{cl}}=\frac1{\sqrt{24}}\tilde g_{\mathrm{cl}} \\
\displaystyle g_{2\mathrm{cl}}=\frac1{\sqrt{8}}\tilde g_{\mathrm{cl}} \\
\displaystyle g_{3\mathrm{cl}}=\frac1{\sqrt{3}}\tilde g_{\mathrm{cl}} \\
\ear\eq}

(If the Dimension-Coupling Hypothesis is correct, the question of coupling constant unification will, therefore, be settled, and the proton will remain stable.  A related discussion of the Standard Model couplings in GUTs is given in [GQW].)

Let us now reformulate the meaning of the $\tilde g_\bk$-ratios in consistency with the Dimen\-sion-Coupling Hypothesis.  We relate the {\em classical} couplings to the {\em physical} ones through
\bq g_{\bk \mathrm{ph}}^2=g_{\bk \mathrm{cl}}^2(1+\Delta_\bk). \eq
Then we have
\bq 
\frac{\dim_\bi}{\dim_\bk}\frac{g_{\bk \mathrm{cl}}^2}{g_{\bi \mathrm{cl}}^2}=\frac{\tilde g_{\bk \mathrm{cl}}^2}{\tilde g_{\bi \mathrm{cl}}^2}
=1,
\eq
while the ratios which were denoted
$\displaystyle\frac{\tilde g_{\bk}^2}{\tilde g_{\bi}^2}$ above, now become, more precisely,
\bq 
\frac{\dim_\bi}{\dim_\bk}\frac{g_{\bk \mathrm{ph}}^2}{g_{\bi \mathrm{ph}}^2}
=\frac{g_{\bi \mathrm{cl}}^2}{g_{\bk \mathrm{cl}}^2}
\frac{g_{\bk \mathrm{ph}}^2}{g_{\bi \mathrm{ph}}^2}
=\frac{1+\Delta_\bk}{1+\Delta_\bi}.
\eq
Thus, equations (12), (14), and (15) indicate the progressive antiscreening effect due to vacuum polarization: the larger the gauge group, the stronger the antiscreening effect [W].  The relationship between $g_{\bk \mathrm{ph}}$ and $g_{\bk \mathrm{cl}}$ is reminiscent of that between the renormalized coupling $g_{\bk \mathrm{R}}$ and the bare
coupling $g_{\bk \mathrm{b}}$ in renormalized perturbation theory. In fact, we take renormalized $g_{\bk \mathrm{R}}$ as our physical $g_{\bk \mathrm{ph}}$. Since our approach does not involve any recourse to perturbation theory we hope to be able to avoid tying our classical $g_{\bk \mathrm{cl}}$ to the bare $g_{\bk \mathrm{b}}$ and infinite renormalization constants. It seems more natural to expect the \mbox{$\Delta_\bk$} to be small 
(i.e.~$|\Delta_\bk|<1$), than having to identify \mbox{$1+\Delta_\bk$} with infinite renormalization constants.  Renormalized perturbation theory implies that any fixed classical value entering the bare Lagrangian may be arbitrarily changed in the quantum theory. In a comparable case, the experimentally observed vanishing of the 
QCD $\theta$-angle is not in agreement with the theoretical prediction that it should assume an arbitrary value [J]. This, together with the fundamental non-perturbative nature of reality, as well as the inherent circularity of the argument concerning the coupling -- which, after all, is the parameter of the perturbative expansion -- seems to justify an approach that does not outright discard the possibility of a finite and definite classical coupling. The crucial element entering our hypothesis seems more suggestive and natural to us than a claim that no classical, geometric information contained in the classical (or bare) couplings could possibly survive renormalization.

\subsection{Note on $\alpha$}
In principle, for spin representations an argument analogous to the one for gauge group representations applies, even though the corresponding dimension factors cancel for
the ratios of coupling constants.  They need to be taken into account, however, if we want to discuss the values of coupling constants rather than just their ratios.

While the analog of the trace-dimension argument for the Lorentz group requires a careful discussion which we don't give here, it can be argued that a factor 3 (rather than 4) will be produced as the dimension of the adjoint representation of spin $\SU(2)$.  This inspires us to set $\tilde g_{\mathrm{cl}} = \sqrt3\,\hat g_{\mathrm{cl}}$, from which we infer

\bq
\alpha_{\mathrm{cl}} = \frac{e_{\mathrm{cl}}^2}{4\pi} = \frac1{4\pi}
\left(\frac1{g_{\mathrm{1cl}}^2}+\frac1{g_{\mathrm{2cl}}^2}\right)^{-1}
= \frac1{4\pi}
\left(\frac{24 + 8}{\tilde g_{\mathrm{cl}}^2}\right)^{-1}
= \frac{3\hat g_{\mathrm{cl}}^2}{32\times 4\pi} 
\approx \frac{\hat g_{\mathrm{cl}}^2}{134}, \eq

and, thereby,

\bq
1+\Delta_e = \frac{e_{\mathrm{ph}}^2}{e_{\mathrm{cl}}^2} 
= \frac{\alpha_{\mathrm{ph}}}{\alpha_{\mathrm{cl}}}
\approx \frac{134}{137}\times\frac1{\hat g_{\mathrm{cl}}^2}
\approx 0.978\times\frac1{\hat g_{\mathrm{cl}}^2}
= (1 - 0.022)\times\frac1{\hat g_{\mathrm{cl}}^2}. \eq

Since screening implies $\;\Delta_e < 0,$ this allows for the possibility that

\bq \hat g_{\mathrm{cl}} = 1. \eq

\newpage

\section*{\normalsize References}

\renewcommand{\arraystretch}{1}
\begin{tabular}{ll}
\mbox{[\enspace]} & Particle Data Group, D.~E.~Groom {\em et al.}, Eur. Phys. J. C {\bf 15}, 1 (2000)  \\
& \\
\mbox{[GQW]} & H.~Georgi, H.~Quinn, and S.~Weinberg, Phys. Rev. Lett. {\bf 33}, 451 (1974) \\
& \\
\mbox{[J]} & R.~Jackiw, ``Topological Investgations of Quantized Gauge Theories'', in \\
& S.~Treiman, R.~Jackiw, B.~Zumino, and E.~Witten, {\em Current Algebra and} \\
& {\em Anomalies} (World Scientific/Princeton University Press, Singapore/ \\
& Princeton NJ 1985), p286 \\
& \\
\mbox{[W]} & F.~Wilczek, ``Beyond the Standard Model: An Answer and Twenty \\
& Questions'', hep-ph/9802400, p10 \\
\end{tabular}

\end{document}